\documentclass[journal]{IEEEtran}
\newtheorem{definition}{Definition}
\usepackage{url}
\usepackage{cite}
\usepackage{latexsym}
\usepackage{amsmath}
\usepackage{amstext,amsfonts,epsf,epsfig,pifont}
\usepackage{graphics,graphicx,bm,amsmath,multicol}

\newlength{\intwidth}

\def\Xint#1{\mathchoice
{\XXint\displaystyle\textstyle{#1}}%
{\XXint\textstyle\scriptstyle{#1}}%
{\XXint\scriptstyle\scriptscriptstyle{#1}}%
{\XXint\scriptscriptstyle\scriptscriptstyle{#1}}%
\!\int}
\def\XXint#1#2#3{{\setbox0=\hbox{$#1{#2#3}{\int}$}
\vcenter{\hbox{$#2#3$}}\kern-.5\wd0}}

\def\dashint{\Xint-}

\newcommand{\degs}{\ensuremath{^{\circ}}}

\newcommand{\seq}{\!\!\!=\!\!\!}

\newcommand{\bu}{\ensuremath{\mathbf{u}}}

\newcommand{\bx}{\ensuremath{\mathbf{x}}}
\newcommand{\by}{\ensuremath{\mathbf{y}}}

\newcommand{\bJ}{\ensuremath{\mathbf{J}}}

\newcommand{\bT}{\ensuremath{\mathbf{T}}}
\newcommand{\bU}{\ensuremath{\mathbf{U}}}

\newcommand{\bX}{\ensuremath{\mathbf{X}}}
\newcommand{\bY}{\ensuremath{\mathbf{Y}}}

\newcommand{\calA}{\ensuremath{\mathcal{A}}}

\newcommand{\calE}{\ensuremath{\mathcal{E}}}

\newcommand{\sgn}{\ensuremath{{\mathrm{sgn}}}}
\begin{document}

\noindent \emph{The following statements are placed here in accordance with the copyright policy of the Institute of Electrical and Electronics Engineers, Inc., available online at}
\url{http://www.ieee.org/web/publications/rights/policies.html}.\\

\noindent
Lilly, J. M., \&  Olhede, S. C. (2010). Bivariate instantaneous \indent frequency and bandwidth. \emph{IEEE Transactions on Signal}\\\indent \emph{Processing}, \textbf{58} (2), 591--603.\\

\noindent This is a preprint version.  The definitive version is available from the IEEE or from the first author's web site, \url{http://www.jmlilly.net}.\\

\noindent \copyright 2010 IEEE. Personal use of this material is permitted. However, permission to reprint/republish this material for advertising or promotional purposes or for creating new collective works for resale or redistribution to servers or lists, or to reuse any copyrighted component of this work in other works must be obtained from the IEEE.\\

\newpage

\title{Bivariate Instantaneous Frequency and Bandwidth}
\author{Jonathan~M.~Lilly,~\IEEEmembership{Member,~IEEE,}
and Sofia~C.~Olhede,~\IEEEmembership{Member,~IEEE}
\thanks{Manuscript submitted \today.  The work of J. M. Lilly was supported  by award \#0526297 from the Physical Oceanography program of the United States National Science Foundation. A collaboration visit by S.~C.~Olhede to Earth and Space Research in the summer of 2006 was funded by the Imperial College Trust.}
\thanks{J.~M.~Lilly is with Earth and Space Research, 2101 Fourth Ave., Suite 1310, Seattle, WA 98121, USA (e-mail: lilly@esr.org).}
\thanks{S.~C.~Olhede is with the Department of Statistical Science, University College London, Gower Street,
London WC1 E6BT, UK (e-mail: s.olhede@ucl.ac.uk).}}

\markboth{IEEE Transactions on Signal Processing}{Lilly \& Olhede: Bivariate Bandwidth}

\maketitle
\begin{abstract}  The generalizations of instantaneous frequency and instantaneous bandwidth to a bivariate signal are derived.  These are uniquely defined  whether the signal is represented as a pair of real-valued signals, or as one analytic and one anti-analytic signal.  A nonstationary but oscillatory bivariate signal has a natural representation as an ellipse whose properties evolve in time, and this representation provides a simple geometric interpretation for the bivariate instantaneous moments.  The bivariate bandwidth is shown to consist of three terms measuring the degree of instability of the time-varying ellipse: amplitude modulation with fixed eccentricity, eccentricity modulation, and orientation modulation or precession. An application to the analysis of data from a free-drifting oceanographic float is presented and discussed.
\end{abstract}

\IEEEpeerreviewmaketitle

\section{Introduction}

\IEEEPARstart{T}HE representation of a nonstationary real-valued signal as an amplitude- and frequency-modulated oscillation has proven to be a powerful tool for univariate signal analysis.  With the analytic signal \cite{vakman77-spu} as the foundation, the instantaneous frequency \cite{gabor46-piee,boashash92a-ieee} and instantaneous bandwidth \cite{cohen} are time-varying quantities which measure, respectively, the local frequency content of the signal and its lowest-order local departure from a uniform sinusoidal oscillation
\cite[p.~1]{lilly09-itit}. The instantaneous quantities are intimately connected with the first two moments of the spectrum of the analytic signal, and indeed can be shown to
decompose the corresponding frequency-domain moments across time \cite[p.~15--16]{cohen}.

An alternative approach to obtaining a time-varying description of signal properties is to define the local moments of some time-frequency distribution.  These quantities are found by marginalizing a weighted time-frequency distribution, see \cite[p.~119--120]{cohen}, that is, by calculating conditional moments in which one treats a time-frequency distribution like a bivariate probability density function. For the Wigner distribution the first conditional moment in frequency is exactly the instantaneous frequency of the analytic signal, and the conditional spread in frequency is one-half the instantaneous bandwidth plus a second curvature term \cite[p.~120]{cohen}. Like \cite[p.~15--16]{cohen} we refer to the time-varying function that integrates to a global moment as an ``instantaneous moment'', and to the conditional moment of a time-frequency distribution as a ``local quantity'' \cite[p.~64--65]{cohen}. In this paper we focus on the interpretation of the instantaneous moments, defined via the analytic signal, and on their relationships with spectral quantities.

In the past few years there has been a great deal of interest in nonstationary bivariate signals, that is, nonstationary signals consisting of a \emph{pair} of values at each time\footnote{An image, by contrast, consists of a signal value which is a function of two variables.}.  A variety of methods have been proposed for their analysis, including: examining \emph{pairs} of time-frequency representations \cite{schreier03b-itsp,schreier08b-itsp}, i.e. the dual frequency spectrum, the Rihaczek distribution, and relatives; creating suitable coherences based on these objects \cite{hindberg07-itsp}; a bivariate version of the empirical mode decomposition \cite{rilling07-ispl}; local analysis of wavelet time-scale ellipse properties \cite{diallo06-geo}; and an extension of wavelet ridge analysis for modulated oscillatory signals
\cite{delprat92-itit,mallat} to the bivariate case \cite{lilly06-npg}.  The  statistical and kinematic properties of stationary bivariate signals have also been recently revisited in a number of studies \cite{schreier03a-itsp,rubin-delanchy08-itsp,medkour08-itsp,walden08-prsla}.

\begin{figure*}[t]
\begin{center}
\includegraphics[width=6in,angle=0]{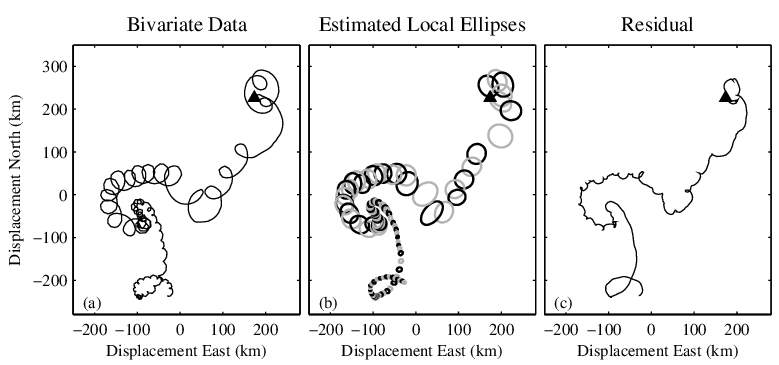}
\end{center}
            \caption{As an example of a bivariate time series, panel (a) shows the trajectory of a freely drifting oceanographic float as it follows the ocean currents in the eastern subtropical Atlantic. The original data (a) is decomposed into a modulated bivariate oscillation (b) plus a residual (c).  In (b), the modulated oscillation is represented by ``snapshots'' at different times, plotted following the instrument location, and black and gray ellipses are alternated for clarity.  The time interval between snapshots varies in time, and is equal to twice the estimated local period of oscillation. The beginning of the record is marked by a triangle in each plot. The origin of these plots is at 22.25\degs N, 25.10\degs W.}\label{bandwidth_looper}
\end{figure*}

As an example of the importance of bivariate nonstationary signals, Fig.~\ref{bandwidth_looper}a shows the trajectory of a freely drifting oceanographic instrument \cite{richardson89-jpo} as it follows the ocean circulation for hundreds of kilometers. Acoustically-tracked subsurface ``floats'' such as this one \cite{rossby86-jaot} are a prominent data source for studying the physics of the turbulent ocean.  There exist hundreds of such records from all over the world, and their analysis constitutes an important and active branch of oceanographic research.  Panels (b) and (c) in this figure show the estimated oscillatory signal and background using the wavelet ridge method of \cite{lilly06-npg}.  The estimated signal, a modulated \emph{bivariate} oscillation, is represented in Fig.~\ref{bandwidth_looper}b as a sequence of ellipses plotted at different times.  The analysis of this data will illustrate the great utility of extending the concept of instantaneous moments to the bivariate case.

The goal of this paper is to extend the main tools for analyzing univariate modulated oscillations---the instantaneous moments---to the bivariate case.  A key ingredient for the analysis of a multivariate signal is the specification of a suitable structural model, the purpose of which is to condense the information from a set of disparate signals into a smaller number of relevant and intuitive parameters.  Work thirty years ago in the oceanographic literature \cite{gonella72-dsr,mooers73-dsr,calman78a-jpo,calman78b-jpo,hayashi79-jas} demonstrated the utility of considering a stationary bivariate signal to be composed of a random ellipse at each frequency.  Here the recently introduced notion of a \emph{modulated elliptical signal} \cite{lilly06-npg} will be used as the foundation for our analysis.

The structure of the paper is as follows. The univariate instantaneous moments are reviewed in Section~\ref{section:background}.  Section~\ref{section:oscillations} introduces the modulated elliptical signal, and expressions for \emph{pairs} of instantaneous moments are derived.  This suggests, for a multivariate signal, defining a \emph{joint instantaneous frequency} and a \emph{joint instantaneous bandwidth} which integrate, respectively, to the frequency and bandwidth of the average of the individual Fourier spectra.  The joint instantaneous moments are derived for general multivariate signals in Section~\ref{section:multivariate}, and the bivariate case is then examined in detail. These quantities are shown to be independent of unitary transformations on the analytic signals, including in particular real rotations of the coordinate axes.  An application to the oceanographic data of Fig.~\ref{bandwidth_looper} is presented in Section~\ref{application}, and the paper concludes with a discussion.

The most important results of this paper concern the instantaneous bandwidth, a fundamental quantity which accounts, together with variations of the instantaneous frequency, for the spread of the Fourier spectrum about its mean \cite{cohen}.  The correct generalization of the univariate bandwidth to the multivariate case has a surprising but intuitive form.  In particular, the bivariate bandwidth admits an elegant geometric interpretation as a fundamental measure of the degree of instability of a time-varying ellipse.  It consists of three terms: root-mean-square amplitude modulation, eccentricity modulation or distortion, and orientation modulation or precession, all of which contribute to the Fourier bandwidth.

All data, numerical algorithms, and scripts for analysis and figure generation are distributed as a part of a freely available package of Matlab routines. This package, called JLAB, is available at the first author's website,
\url{http://www.jmlilly.net}.

\section{Background}

This section gives a compact review of the theory of univariate instantaneous moments, drawing in particular from \cite{cohen}, and gives definitions which will be used throughout the paper.

\subsection{The Analytic Signal}\label{section:background}
A powerful model for a real-valued nonstationary signal $x(t)$, assumed deterministic and square-integrable herein, is the modulated oscillation
\begin{equation}
x(t) =a_x(t)\cos \phi_x(t)\label{realmodulated}
\end{equation}
where $a_x(t)\ge 0$ and $\phi_x(t)$ are here defined to be particular unique functions called the instantaneous \emph{canonical amplitude} and \emph{canonical phase}, respectively \cite{picinbono83-adt,picinbono97-itsp}.

The canonical pair $[a_x(t),\phi_x(t)]$ is defined in terms of the analytic signal $x_+(t)$, itself defined as \cite{gabor46-piee,boashash92a-ieee}
\begin{eqnarray}
x_+(t)&\equiv & 2\calA [x](t)\equiv x(t)+i\frac{1}{\pi}\,\dashint_{-\infty}^\infty\frac{ x(u)}{ t-u}\,d u
\label{analyticsignal}
\end{eqnarray}
where ``$\dashint$'' is the Cauchy principal value integral and  $\calA$ is called the \emph{analytic operator}.  This creates a unique complex-valued object $x_+(t)$ from any real-valued signal $x(t)$, and permits $x(t)$ to be associated with a unique amplitude and phase through
\begin{equation}
a_x(t)e^ {i\phi_x (t)} \equiv  x_+(t),
\end{equation}
with the canonical pair $[a_x(t),\phi_x(t)]$ defined by
\begin{eqnarray}
a_x(t)&=&\left|x_+(t)\right|\\
\phi_x(t)&=&\tan^{-1}
\left(\frac{\Im\{x_+(t)\}}{\Re\{x_+(t)\}} \right),
\quad a_x(t)\neq 0
\end{eqnarray}
where ``$\Re$''  denotes the real part and ``$\Im$''  denotes the imaginary part.
Note that at isolated points where $a_x(t)=0$, the value of the phase is typically defined by continuity \cite{picinbono97-itsp}. The original signal is then recovered from the analytic signal via $x(t)=\Re\left\{x_+(t)\right\}$.

The action of the analytic operator is more clear in the frequency domain.  $X(\omega)$ is the Fourier transform of $x(t)$,
\begin{eqnarray}
x(t)&= & \frac{1}{2\pi}
\int_{-\infty}^\infty X(\omega)\,e^{i \omega t}\,d \omega
\label{inverseFourier}
\end{eqnarray}
and likewise  $X_+(\omega)$ is the Fourier transform of $x_+(t)$.  The time-domain operator in (\ref{analyticsignal}) becomes in the frequency domain simply \cite{poletti97-itsp}
\begin{equation}
X_+(\omega)\equiv 2U(\omega) X(\omega)\label{analyticFourier}
\end{equation}
where $U(\omega)$ is the unit step function. Thus in the frequency domain, the analytic signal is formed by doubling the amplitudes of the Fourier coefficients at all positive frequencies, while setting those of all negative frequencies to zero.

It should be recognized that, in general, the choice of an amplitude and phase pair of functions in (\ref{realmodulated}) for a given signal $x(t)$ is not unique  (see e.g. \cite{loughlin97-ispl}).  However, the canonical pair is of fundamental importance because it is an objective and well-established method of assigning an amplitude and phase to an observed oscillatory signal.  The properties of the canonical pair have been thoroughly investigated by other authors; see the discussion and references in \cite{boashash92a-ieee} and \cite{picinbono97-itsp}.  One attractive property of the analytic signal is that if one constructs an $x(t)=a(t)e^{i \phi(t)}$ based on some amplitude $a(t)$ and phase $\phi(t)$, where the frequency-domain support of $a(t)$ and $e^{i \phi(t)}$ have no overlap, then the analytic signal will recover the original amplitude and phase [Bedrosian's theorem,~\citen{bedrosian63-ire}].  This would be case if, for example, $a(t)$ is low-frequency signal while $e^{i \phi(t)}$ is a high-frequency signal.  One should note, however, that there exist cases for which the canonical pair associated with the analytic signal does not yield useful information; an important example is that of a time series consisting of an aggregation of oscillatory signals at different frequencies.

\subsection{Global and Instantaneous Moments}
Using the analytic signal, one may describe both the local and the global behavior of any real-valued signal $x(t)$ as a modulated oscillation. A global description is afforded by the frequency-domain moments of the spectrum of the analytic signal.  The \emph{global mean frequency} and \emph{global second central moment} of $x(t)$ are defined, respectively, by
\begin{eqnarray}
\overline \omega_{x}  &\equiv &\frac{1}{2\pi \calE_x}\int_{0} ^\infty \omega\left| X_+(\omega)\right|^ 2d\omega
\label{globalfrequency}\\
\overline \sigma_x^2  &\equiv& \frac{1}{2\pi \calE_x}\int_{0} ^\infty \left (\omega-\overline  \omega_x\right)^2\left| X_+(\omega)\right|^ 2d\omega
\label{nthmoment}
\end{eqnarray}
where
\begin{equation}
 \calE_x\equiv \frac{1}{2\pi}\int_{0} ^\infty  \left|X_+(\omega)\right|^ 2d\omega=\int_{-\infty} ^\infty \left|x_+ (t)\right|^ 2d t
\end{equation}
is total energy of the analytic part of the signal.   The second central moment $\overline \sigma_x^2$ characterizes the spread of the spectrum about the global mean frequency; thus $\overline \sigma_x$ is also known as the \emph{bandwidth}.

It would be greatly informative to relate the global moments $\overline\omega_x$ and $\overline \sigma_x$ to the time evolution of the signal.  This leads to the notion of \emph{instantaneous moments}, time-varying quantities which recover the global moments through integrals of the form\cite{cohen,loughlin00-ispl}
\begin{eqnarray}
\overline \omega_{x} & =  &\calE_x^ {-1}\int_{-\infty}^\infty a_x^ 2(t) \,\omega_{x}(t)\,d t\label{instantaneousfrequency}\\
\overline \sigma_x^2 & =  &\calE_x^ {-1}\int_{-\infty}^\infty a_x^ 2(t) \,\sigma_x^2(t)\,d t.\label{instantaneousCentralmoment}
\end{eqnarray}
It is a fundamental result that the derivative of the canonical phase
\begin{eqnarray}
\omega_x(t)&\equiv&\frac{d}{d t}\,\phi_x(t) =\Im\left\{\frac{d}{d t}\ln x_+(t)\right\}\label{instantaneousfrequencydefinition}
\end{eqnarray}
satisfies (\ref{instantaneousfrequency}), and is termed therefore the \emph{instantaneous frequency} \cite{gabor46-piee,boashash92a-ieee}. It is important to point out that the instantaneous moments $\omega_{x}(t)$ and $\sigma_x^2(t)$ are not themselves time-domain moments; rather, they are instantaneous \emph{contributions to} the global spectral moments. \footnote{We point out that while the instantaneous frequency $\omega_x(t)$ may concur with the local frequency found by calculating a conditional moment of a chosen time-frequency distribution \cite[p.~119]{cohen}, this correspondence between instantaneous moments and local moments does not extend to higher orders for the Wigner distribution \cite[p.~120]{cohen}.}

The instantaneous second central moment $\sigma_x^2(t)$ is not uniquely defined by the integral (\ref{instantaneousCentralmoment}), because more than one function of time may be shown to integrate to the corresponding global moment.
(The same is true for $\omega_x(t)$ as we could add any function that when multiplied by $a_x^2(t)$ integrates to zero.)  However, the constraint that $\sigma_x^2(t)$ be nonnegative definite, like $\overline \sigma_x^ 2$, leads to the unique definition \cite{cohen}
\begin{equation}
\sigma_x^2(t)\equiv  \frac{\left|\frac{d }{dt} \,x_+(t)-i\overline \omega_xx_+(t) \right|^2}{|x_+(t)|^2}.\label{secondcentralinstantaneous}
\end{equation}
The square root of the instantaneous second central moment so defined, $\sigma_x(t)$, then has an intuitive interpretation as the average spread of the frequency content of the signal at each point in time about the global mean frequency $\overline \omega_x$.

It has been shown by \cite{cohen} that the fractional rate of amplitude modulation
\begin{equation}
\upsilon_x(t)\equiv \frac{d\ln a_x(t)}{d t}=\Re\left\{ \frac{d}{d t}\,\ln x_+(t)\right\}\label{instantaneousbandwidthdefinition}
\end{equation}
plays an important role within the instantaneous second central moment. After a simplification, (\ref{secondcentralinstantaneous}) becomes
\begin{equation}
\sigma_x^2(t)=\left[\omega_x(t)-\overline \omega_x\right]^ 2+\upsilon_x^2(t)\label{secondcentralinstantaneousdefinition}
\end{equation}
from which it is clear that variations of the instantaneous frequency $\omega_x(t)$ about the global mean frequency $\overline \omega_x$ do not account for the entirety of the global second central moment $\overline \sigma_x^2$.  The remainder is accounted for by $\upsilon_x^ 2(t)$, and since $\overline \sigma_x$ is known as the global bandwidth, $\upsilon_x(t)$ is called the \emph{instantaneous bandwidth} \cite{cohen,cohen89-ieee,cohen88-spie}.

\section{The Modulated Elliptical Signal}\label{section:oscillations}
This section introduces a model for a modulated bivariate oscillation, extending and simplifying the development in \cite{lilly06-npg}.

\subsection{Modulated Ellipse Model}

Bivariate signals will be represented both in the complex-valued form  $z(t)\equiv x(t)+i  y(t)$, and in the vector form $\bx(t)\equiv \begin{bmatrix} x(t)&  y(t)\end{bmatrix}^T$, with $x(t)$ and $y(t)$ both real-valued and where ``$T$'' denotes the matrix transpose.  Shortly we will take the analytic part of $\bx(t)$, to give $\bx_+(t)\equiv 2\calA[\bx](t)$.  Note that one may recover the complex-valued signal $z(t)$ in terms of the pair of analytic signals as
\begin{equation}
z(t)= \Re\left\{x_+(t)\right\}
        + i  \Re\left\{y_+(t)\right\}. \label{complexbivariateform}
\end{equation}
The complex-valued signal $z(t)$ corresponding to a bivariate real-valued vector does not, in general, have its imaginary part and real part being Hilbert transforms of each other, but is formed from two separately observed signals, rather than from a single signal.

\begin{figure}[t]
\begin{center}
\includegraphics[height=3.5in,angle=0]{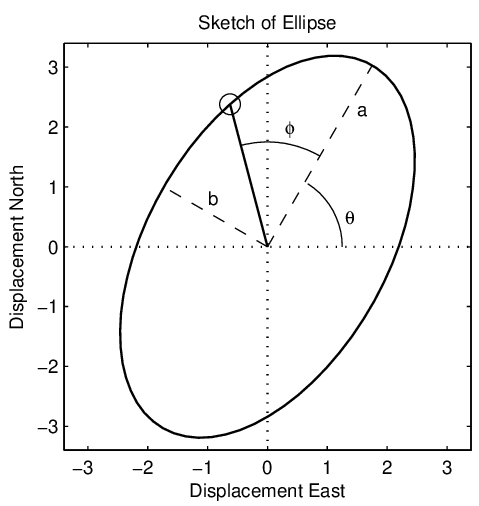}
\end{center}
            \caption{A sketch of the parameters in the modulated elliptical  signal, as described in the text.}\label{ellipsesketch}
\end{figure}

Our starting point for describing bivariate oscillations will be the \emph{modulated elliptical signal}  \cite{lilly06-npg}
\begin{equation}
\label{ellipseeqn}
z(t)  \equiv e^{i\theta(t)}\left\{
        a(t)\cos\phi(t)+i  b(t)\sin\phi(t)\right\}
\end{equation}
which represents $z(t)$ as the position traced out on the $[x,y]$ plane by a hypothetical particle orbiting a time-varying ellipse; see the sketch in Fig.~\ref{ellipsesketch}.  The two angles $\theta(t)$ and $\phi(t)$ are defined on the principal interval $(-\pi,\pi)$, while $a(t)\ge 0$; $b(t)$ may take on either sign for reasons to be see shortly.

The ellipse has instantaneous semi-major and semi-minor axes $a(t)$ and $|b(t)|$, and an instantaneous orientation of the major axis with respect to the $x$-axis given by $\theta(t)$. The angle $\phi(t)$, called the \emph{orbital phase}, determines the instantaneous position of the particle around the ellipse perimeter with respect to the major axis.  While $a(t)$ is nonnegative, the sign of $b(t)$ is chosen to reflect the direction of circulation around the ellipse.  Note that the modulated univariate signal (\ref{realmodulated}) is included as the special case $b(t)=0$, $\theta(t)=0$.  Thus we can view the modulated elliptical signal as an extension of the amplitude/frequency modulated signal to the bivariate case.

The meaning of the ``instantaneous'' ellipse properties referred to in the previous paragraph is more clear if we separate the ellipse phase from the other variables. Introduce
\begin{multline}
\label{ellipseeqnseparated}
z(t, t')  \equiv e^{i\theta(t)}\left\{
        a(t)\cos\left[\phi(t)+\omega_\phi(t)t'\right]\right.\\\left.+i  b(t)\sin\left[\phi(t)+\omega_\phi(t)t'\right]\right\}
\end{multline}
where $t'$ is seen as a ``local'' time.  With $t$ fixed, $z(t,t')$ continually traces out the perimeter of a ``frozen'' ellipse as $t'$ varies; the period of orbiting the ellipse is $2\pi/\omega_\phi(t)$.  It is these frozen ellipses which are plotted in Fig.~\ref{bandwidth_looper}b for the estimated modulated elliptical signal corresponding to the time series in Fig.~\ref{bandwidth_looper}a. Note that $z(t)$ itself will not in general trace out an ellipse if the ellipse geometry varies in time. However, if the parameters of the ellipse geometry $a(t)$, $b(t)$, and  $\theta(t)$ are slowly varying with respect to the phase $\phi(t)$, then $z(t)$ will \emph{approximate} an ellipse.

\subsection{Analytic Signal Pairs}

The parameters of the modulated elliptical signal are made unique for a given pair of real-valued signals by referring to a \emph{pair} of analytic signals, just as the univariate modulated signal model is made unique by referring to a single analytic signal.  In vector notation the modulated elliptical signal (\ref{ellipseeqn}) becomes
\begin{equation}
\begin{bmatrix}x(t)\\
  y(t)
   \end{bmatrix}=\bJ\left(\theta(t)\right)\Re\left\{e^{i\phi(t)}\begin{bmatrix}a(t)\\
 -ib(t)
   \end{bmatrix}\right\}\label{realellipsematrixform}
\end{equation}
where
\begin{equation}
\bJ(\theta) =
    \begin{bmatrix}\cos\theta&-\sin\theta \\
   \sin\theta & \cos\theta
   \end{bmatrix}
\end{equation}
is the counterclockwise rotation matrix.
This permits us to introduce a new useful representation of the signal.
\begin{definition}{The Ellipse Parameters}\\
We define the \emph{canonical} ellipse parameters $a(t)$,
$b(t)$, $\theta(t)$ and $\phi(t)$ in terms of the analytic portions of  $x(t)$ and $y(t)$ implicitly by the equality
\begin{equation}
e^{i\phi(t)}\bJ\left(\theta(t)\right)\begin{bmatrix}a(t)\\
 -ib(t)
   \end{bmatrix}\equiv \begin{bmatrix}x_+(t)\\
  y_+(t)
   \end{bmatrix}=\begin{bmatrix} 2\calA[x](t)\\
   2\calA[y](t)
   \end{bmatrix}.\label{ellipsematrixform}
\end{equation}
Explicit expressions for $a(t)$, $b(t)$,  $\theta(t)$, and  $\phi(t)$ in terms of $x_+(t)$ and $y_+(t)$ will be given shortly. \end{definition}

We emphasize that while the quantities $a(t)$, $b(t)$, $\theta(t)$, and $\phi(t)$ appearing in  (\ref{realellipsematrixform}) could be specified arbitrarily, we have chosen to define them in terms of $x_+(t)$ and $y_+(t)$ as in (\ref{ellipsematrixform}).  This precisely parallels the assignment of a canonical amplitude and phase to a univariate signal using its analytic part, as discussed in Section~\ref{section:background}. The canonical ellipse parameters are therefore unique properties of a given bivariate signal.

While any bivariate signal can be represented as a modulated ellipse, just as any univariate signal can be represented as a modulated oscillation via the analytic signal, it is not always sensible to do so.  If one begins with a pair of real-valued signals $x(t)$ and $y(t)$ which are, say, finite samples of a noise process, the ellipse parameters would likely not yield any sensible information; this would inform you that the modulated ellipse representation is not appropriate.  Another situation in which the modulated ellipse representation would not be appropriate is for signals consisting of aggregations of several oscillatory components \cite{loughlin97-ispl,oliveira98-ieee}, as discussed in \cite[p.~527]{boashash92a-ieee} or \cite[p.~626]{born}; we do not consider such multi-component signals in this paper.

One may also associate a second pair of analytic signals with $z(t)$, namely\begin{equation}
\left[\begin{array}{c}
 z_+(t)\\z_-(t)
\end{array}\right]\equiv \begin{bmatrix} \calA[z](t)\\
   \calA[z^*](t)
   \end{bmatrix}\label{pnanalytic}
\end{equation}
the components of which are the analytic parts of $z(t)$ and of its complex conjugate. Then
\begin{eqnarray}
z(t)&= & z_+(t)+z_-^*(t) \label{complexbivariateform2}
\end{eqnarray}
decomposes $z(t)$ into counterclockwise and clockwise rotating contributions, respectively.\footnote{The factor of two difference in defining analytic versions of real-valued and complex-valued signals, i.e.  the last equality in (\ref{ellipsematrixform}) versus (\ref{pnanalytic}), prevents factors of two from appearing in the decomposition equations (\ref{complexbivariateform}) and (\ref{complexbivariateform2}).}  Note that for convenience we have defined $z_-(t)$ to be an \emph{analytic}, rather than an anti-analytic, signal.  Making use of the unitary matrix
\begin{equation}
 \bT\equiv \frac{1}{ \sqrt{2}} \left[\begin{array}{cc}
1 &i\\1 &-i
\end{array}\right]
\end{equation}
one finds
\begin{equation}
\left[\begin{array}{c}
 z_+(t)\\z_-(t)
\end{array}\right]=\frac{1}{\sqrt{ 2}} \bT\left[\begin{array}{c}
 x_+(t)\\y_+(t)
\end{array}\right]\label{rotarytoCartesian}
\end{equation}
as the relationship between the two different pairs of analytic signals.  In order to distinguish between these two pairs, we refer to $x_+(t)$ and $y_+(t)$ as the \emph{Cartesian pair} of analytic signals, and to $z_+(t)$ and $z_-(t)$ as the \emph{rotary pair} (borrowing a term from the oceanographic community \cite{gonella72-dsr,mooers73-dsr}).

The ellipse parameters have simple expressions in terms of the amplitudes and phases of the rotary analytic signals. Letting
\begin{eqnarray}
 z_+(t)&=  &a_+(t)e^{i\phi_+(t)}\label{zplus}\\
 z_-(t)&  =&a_-(t)e^{i\phi_-(t)}\label{zminus}
\end{eqnarray}
and comparing the rotary decomposition (\ref{complexbivariateform2}) with $\bT$ times the matrix form of the modulated ellipse model (\ref{ellipsematrixform}), one finds \cite{lilly06-npg}
\begin{eqnarray}
a(t)&=& a_+(t)+a_-(t)\label{majoraxis}\\
b(t)&=& a_+(t)-a_-(t)\label{minoraxis}\\\phi(t)&=& \left[\phi_+(t)+\phi_-(t)\right]/2\label{phidefinition}\\
\theta(t)&=& \left[\phi_+(t)-\phi_-(t)\right]/2\label{thetadefinition}
\end{eqnarray}
as expressions for the ellipse parameters in terms of the amplitude and phases of the rotary analytic signals.  The ellipse parameters are therefore uniquely defined.  Unique relations of the ellipse parameters to the parameters of the Cartesian analytic signals are then implied by (\ref{rotarytoCartesian}), and are given in Appendix~\ref{section:Cartesian}.

\subsection{Amplitude and Eccentricity}
It is convenient to replace the semi-major and semi-minor axes $a(t)$ and $b(t)$ with the root-mean-square amplitude
 \begin{equation}
\kappa(t)\equiv \label{amplitudedefinition} \sqrt{\frac{a^2 (t)+b^2(t)}{2}}
\end{equation}
together with a relative of the eccentricity \cite{ruddick87-jpo},\cite{lilly06-npg}
\begin{equation}
 \label{eq:lambda}
\lambda(t)  \equiv r_z\frac{a^2(t)-b^2(t)}{a^2(t)+b^2(t)}
\end{equation}
which measures the signal's departure from circularity\footnote{Note that herein we always use the term ``circular'' to describe the shape traced out on the [$x$,$y$] plane by a bivariate signal, rather than in the statistical sense of a circularly symmetric or proper complex-valued signal [e.g. \citen{schreier03a-itsp}].}.  Here the direction of rotation of the ``particle'' around the time-varying ellipse is denoted by
\begin{equation}
r_z\equiv \sgn\left\{b(t)\right\}
\end{equation}
which we assume henceforth to be constant over the time interval of interest.  These definitions give
\begin{eqnarray}
a(t)&=&\kappa(t)\sqrt{1+\left|\lambda(t)\right|}\label{adef}\\
b(t)&=&r_z\kappa(t)\sqrt{1-\left|\lambda(t)\right|}\label{bdef}
\end{eqnarray}
as expressions for the semi-major and (signed) semi-minor axes. Note that $r_z$ is well defined for purely circular signals for which $a(t)=|b(t)|$, whereas $\sgn\left\{\lambda(t)\right\}$ vanishes.

The magnitude of $\lambda(t)$, like the eccentricity $\mathrm{ecc}(t)=\sqrt{1 -b^2(t)/a^2(t)}$,  varies between $|\lambda(t)|=0$ for purely circularly polarized motion and $|\lambda(t)|=1$ for purely linearly polarized motion.  We refer to $|\lambda(t)|$ as the ellipse \emph{linearity} and to  $\lambda(t)$ as the \emph{signed linearity}.  While not in common use as a measure of eccentricity, we prefer $\lambda(t)$ to other such measures since it leads to simple forms for subsequent expressions.

\subsection {Rates of Change}

The rates of change of the ellipse parameters have special interpretations. The time derivative of the orbital phase, termed the \emph{orbital frequency} $\omega_\phi(t)\equiv \frac{d}{d t} \, \phi(t)$, gives the rate at which the particle orbits the ellipse, while the time derivative of the orientation angle is the \emph{precession rate}  $\omega_\theta(t)\equiv \frac{d}{d t} \, \theta(t)$. Amplitude modulation of the ellipse involves a time derivative of $\kappa(t)$, while variation in $\lambda(t)$ means that the shape of the ellipse is distorting with time.  However, the ellipse parameters and their derivatives do not have an immediately evident relationship to the global moments of $z(t)$. Thus, while these are useful local descriptions of joint signal variability, they are not interpretable as instantaneous moments.

\subsection{Instantaneous Moment Pairs}

In Section~\ref{section:background} it was shown that the instantaneous moments of a univariate signal---instantaneous frequency and bandwidth, in particular---provide a powerful description of local signal variability with a direct relationship to the global signal moments.  We aim to identify the analogous quantities for a bivariate signal.

A bivariate oscillatory signal can be equivalently expressed in terms of two \emph{pairs} of analytic signals, the Cartesian pair $x_+(t)$ and $y_+(t)$ or the rotary pair $z_+(t)$ and $z_-(t)$.  Each of these, in turn, can be analyzed by their individual instantaneous moments, using the ideas described in Section~\ref{section:background}. In particular, these two pairs of analytic signals lead to two pairs of instantaneous frequencies $[\omega_x(t)$, $\omega_y(t)$] and [$\omega_+(t)$, $\omega_-(t)$] and two pairs of instantaneous bandwidths [$\upsilon_x(t)$, $\upsilon_y(t)$] and [$\upsilon_+(t)$, $\upsilon_-(t)$], which contribute to two corresponding pairs of global moments.  This approach is, however, unsatisfactory.  Rather than a unified description of signal variability, an analysis based on these pairs of instantaneous moments achieves, in two different ways, a description of two disparate halves of the bivariate signal $z(t)$ considered separately.

The difficulty of a pairwise decomposition is highlighted by considering a range of different values of the linearity $|\lambda(t)|$. Generally speaking, the decomposition of $z(t)$ into the rotary pair of analytic signals is more appropriate when one of $z_+(t)$ and $z_-(t)$ is much larger than then other, e.g. $|z_+(t)|\gg|z_-(t)|$.  On the other hand, when (in some rotated coordinate system) one has $|x_+(t)|\gg|y_+(t)|$, then the Cartesian decomposition is more appropriate.  These two cases correspond to $|\lambda(t)|\approx 0$ and $|\lambda(t)|\approx 1$, respectively.

In between these two cases, or for a signal ranging over both extremes, there is a gray area in which neither decomposition is particularly appropriate.
To work with such signals one approach would be to replace \eqref{rotarytoCartesian} by
\begin{equation}
\bu_+(t;\bU)=\bU(t)\bx_+(t)\label{rotarytoCartesian2}
\end{equation}
and then to choose $\bU(t)$ in order to maximize the amplitude of the first component of $\bu_+(t;\bU)$.  In fact, this is precisely what the ellipse analysis accomplishes with
\begin{equation}
\bU_o (t) \equiv \frac{1}{\kappa(t)}\begin{bmatrix} a(t) & b(t) \\-b(t) & a(t)\end{bmatrix}\begin{bmatrix} 1 & 0 \\0 & i
\end{bmatrix}\bJ^T(\theta(t)).
\end{equation}
The reader may verify that with $\bU_o(t)$ so defined, one has $\bu_+(t;\bU_o)=\kappa(t)e^{i\phi(t)}\times\begin{bmatrix}2 & 0 
\end{bmatrix}^T$.  Thus the ellipse representation obtains the most compact description of the signal over the full range of polarizations.


The instantaneous moments of the analytic signal pairs can be cast in terms of the ellipse parameters, to show how changes in the ellipse geometry are expressed in time variability of the instantaneous moments pairs.  Inverting (\ref{majoraxis}--\ref{thetadefinition}), and utilizing (\ref{adef}) and (\ref{bdef}), leads to
\begin{eqnarray}
\phi_\pm(t)& = &   \phi(t)\pm\theta(t)\label{phiplusminus} \\a_\pm(t)& =&\frac{\kappa(t)}{\sqrt{2}}\sqrt{1\pm r_z\sqrt{1-\lambda^2(t)}} \label{plusminusamplitude}
\end{eqnarray}
giving the parameters of the rotary analytic signals in terms of the ellipse parameters. (In these and subsequent expressions, the signs on the right-hand-side are understood to be chosen to match the signs on the left.)  Note that the amplitudes satisfy  $a_+^ 2(t)+a_-^ 2(t)=\kappa^ 2(t)$, and we may identify  $\kappa^ 2(t)$ as the sum of the instantaneous power of the two rotary signals.  When $r_z>0$, and the ellipse rotates in the positive (counterclockwise) direction, we have $a_+(t)>a_-(t)$ as expected, whereas the opposite is true for negative rotation.  Note that as $|\lambda(t)|$ approaches zero, and the signal becomes nearly circular, the smaller of $a_+(t)$ and $a_-(t)$ approaches zero while the larger approaches $\kappa(t)$.


Differentiating (\ref{phiplusminus}) and also the logarithm of (\ref{plusminusamplitude}) leads to
\begin{eqnarray}
\omega_{\pm}(t)&=&\omega_\phi(t)\pm\omega_\theta(t)\label{freqgreater}\\
\upsilon_\pm(t) 
  &= &  \frac{d\ln\kappa(t)}{d t}\pm r_z\frac{1}{2}\frac{\frac{d}{d t}\sqrt{1-\lambda^2(t)}}{1\pm r_z\sqrt{1-\lambda^2(t)}}\label{bandwidthgreater}
\end{eqnarray}
as expressions for the rotary instantaneous frequencies and bandwidths.  It is useful to note an asymmetry of the rotary bandwidths. For small $|\lambda(t)|$ with $r_z>0$, the denominator of the second term in $\upsilon_-(t)$ is close to zero while that of $\upsilon_+(t)$ is close to unity; the situation is reversed for $r_z<0$.   This shows that for small $|\lambda(t)|$, the bandwidth of the weaker of the two rotary signals is much more sensitive to variations in the degree of eccentricity than is the bandwidth of the stronger signal.

As discussed above, these expressions are useful in the near-circular case $|\lambda(t)|\approx 1$.  The same procedure carried out in terms of the Cartesian analytic signals $x_+(t)$ and $y_+(t)$, presented in Appendix~\ref{section:Cartesian},
leads to more complicated relationships between variations of the ellipse geometry and the instantaneous moments of the two real-valued signals $x(t)$ and $y(t)$.  In particular, we may note in those relationships the explicit dependence on the ellipse orientation $\theta(t)$. This renders the Cartesian instantaneous moments quite problematic for a unified description of the signal variability, since a simple coordinate rotation will cause these quantities to change.

There is therefore a disconnect between the moment-based description of bivariate signal variability, grounded on the analytic signals, and the modulated ellipse model of joint structure. This motivates the development of the next section.

\section{Joint Instantaneous Moments}\label{section:multivariate}

The definitions of instantaneous moments in Section~\ref{section:background}, which are standard, can be extended to accommodate multivariate oscillatory signals.  Let
\begin{equation}
\bx_+(t)\equiv \left[x_{+;1}(t)\,\,x_{+;2}(t)\,\,\ldots \,\,x_{+;N} (t)\right]^T
\end{equation}
be a vector of $N$ analytic signals, and let $\bX_+(\omega)$ be the corresponding frequency-domain vector; these are related by the inverse Fourier transform
\begin{equation}
\bx_+(t)=\frac{1}{2\pi}\int_{-\infty}^\infty e^{i\omega t}\, \bX_+(\omega)\, d\omega.
\end{equation}
Here we begin with the general multivariate case and subsequently examine the bivariate case, $N=2$, in detail.

\subsection{Joint Moments}
Let us say we a have vector-valued signal $\bx(t)$ and are interested in a obtaining unified description of the variability of its components. To this end it is reasonable to define the \emph{joint analytic spectrum} as the normalized average of the spectra of its $N$ components,
\begin{eqnarray}
S_\bx(\omega)&\equiv &\calE_\bx^ {-1}\|\bX_+(\omega)\|^ 2
\end{eqnarray}
where $\|\bx\|\equiv \sqrt{\bx^H\bx}$ is the Euclidean norm of a vector $\bx$, ``$H$'' indicating the conjugate transpose, and where
\begin{equation}
\calE_\bx\equiv \frac{1}{2\pi}\int_{0}^\infty\|\bX_+(\omega)\|^ 2d \omega=\int_{-\infty}^\infty\|\bx_+(t)\|^ 2d t
\end{equation}
is the {\em joint  (total) energy} of the multivariate analytic signal.  One may then define the \emph{joint global mean frequency}
\begin{equation}
\overline \omega_{\bx}  \equiv \frac{1}{2\pi\calE_\bx}\int_{0} ^\infty \omega S_\bx(\omega)\,d\omega\label{multivariateglobalfrequency}
\end{equation}
associated with the joint analytic spectrum, together with
\begin{equation}
\overline \sigma_{\bx}^2  \equiv \frac{1}{2\pi\calE_\bx}\int_{0} ^\infty \left(\omega-\overline\omega_{\bx}\right)^2 S_\bx(\omega)\,d\omega\label{nthmomentzcentral}
\end{equation}
which is the \emph{joint global second central moment}.

The \emph{joint instantaneous frequency}  $\omega_{\bx} (t)$ and \emph{joint instantaneous second central moment} $\sigma_{\bx}^2 (t)$  of $\bx_+(t)$ are then some quantities that decompose the corresponding global moments across time, i.e. which satisfy
\begin{eqnarray}
\overline \omega_{\bx} & = &\calE_\bx^ {-1}\int_{-\infty}^\infty \|\bx_+(t)\|^ 2\,\omega_{\bx} (t)\,d  t\label{multivariateinstantaneousmoment}\\
\overline \sigma_{\bx}^2  & = &\calE_\bx^ {-1}\int_{-\infty}^\infty \|\bx_+(t)\|^ 2\,\sigma_{\bx}^2(t)\,d  t\label{multivariateinstantaneouscentralmoment}
\end{eqnarray}
noting that $\|\bx_+(t)\|^ 2$ is the instantaneous total analytic signal power. Since the univariate instantaneous frequency (\ref{instantaneousfrequencydefinition}) may be rewritten
\begin{equation}
\omega_x(t)\equiv \Im\left\{\frac{d\ln x_+(t)}{d t}\right\}=\frac{\Im\left\{x_+^*(t)\frac{d x_+(t)}{d t}\right\}}{\left|x_+(t)\right|^ 2}
\end{equation}
we define the joint instantaneous frequency according to the same form,
\begin{equation}
\omega_{\bx}(t)\equiv\frac{\Im\left\{\bx_+^H(t)\frac{d\bx_+(t)}{d t}\right\}}{\|\bx_+(t)\|^ 2}\label{multivariatefrequency}
\end{equation}
and find that this does indeed satisfy (\ref{multivariateinstantaneousmoment}). \footnote{It is interesting to note that this definition is consistent with the weighted average instantaneous frequency of \cite{loughlin01-itsp,olhede04-prsla}, defined for single observations of multi-component signals, as opposed to the multivariate signals considered here.}

Likewise, the second instantaneous central moment, which we define as
\begin{equation}
\sigma_{\bx}^2(t)\equiv\label{multivariatesecondcentral} \frac{\left\|\frac{d}{d t}\bx_+(t)-i\overline\omega_\bx\bx_+(t)\right\|^ 2}{\|\bx_+(t)\|^ 2}
\end{equation}
is a nonnegative-definite quantity satisfying (\ref{multivariateinstantaneouscentralmoment}).  This gives the normalized departure of the rate of change of the vector-valued signal from a uniform complex rotation at the constant single frequency $\overline\omega_\bx$. Equations (\ref{multivariatefrequency}) and (\ref{multivariatesecondcentral}) are clearly the natural generalizations of (\ref{instantaneousfrequencydefinition}) and (\ref{secondcentralinstantaneous}) to multivariate analytic signals.

\subsection{Joint Instantaneous  Bandwidth}
Furthermore we may generalize the notion of instantaneous bandwidth to a multivariate signal. We define the \emph{joint instantaneous bandwidth} via its relationship to the joint second central moment
\begin{eqnarray}
\upsilon_{\bx}^2(t) &\equiv & \sigma_{\bx}^2(t)-\left[\omega_\bx(t)-\overline\omega_\bx\right]^ 2 
\end{eqnarray}
by comparison with (\ref{secondcentralinstantaneousdefinition}) for the univariate case. That is,  we \emph{define} the squared instantaneous bandwidth to be that part of the instantaneous second central moment not accounted for by deviations of the instantaneous frequency from the global mean frequency.

This definition of the bandwidth leads, after some manipulation, to
\begin{equation}
\upsilon_{\bx}^2(t) =\frac{\left\|\frac{d}{d t}\bx_+(t)-i \omega_\bx(t)\bx_+(t)\right\|^ 2}{\|\bx_+(t)\|^ 2}\label{twobandwidthdef}
\end{equation}
which is the normalized departure of the rate of change of the vector-valued signal from a uniform complex rotation at a single \emph{time-varying} frequency $\omega_\bx(t)$. For $\bx_+(t)$ a 1-vector consisting of a single analytic signal, $x_+(t)$, (\ref{twobandwidthdef}) becomes
\begin{equation}
\upsilon_{\bx}^2(t) = \frac{\left |\left[\frac{d}{d t}-i \omega_x(t)\right]x_+(t)\right|^ 2}{|x_+(t)|^ 2}\label{instantaneousbandwidthexpression}
=\left|\frac{d}{d t}\ln a_x(t)\right|^ 2=\upsilon_x ^ 2(t)
\end{equation}
and the joint instantaneous bandwidth correctly reduces to the univariate bandwidth defined in (\ref{instantaneousbandwidthdefinition}).

The contributions to the instantaneous second central moment and the instantaneous bandwidth are perhaps more clear if we write (\ref{multivariatesecondcentral}) and (\ref{twobandwidthdef}) out as summations
\begin{eqnarray}
\sigma_{\bx}^2(t) &\seq&\frac{\sum_{n=1}^N a_{n}^2(t)\left\{ \upsilon_{n}^ 2(t)+ \left|\omega_{n}(t)-\overline\omega_\bx\right|^ 2\right\}}{\sum_{n=1}^N a_{n}^2(t)} \\ \upsilon_{\bx}^2(t) &\seq&\frac{\sum_{n=1}^N a_{n}^2(t)\left\{ \upsilon_{n}^ 2(t)+ \left|\omega_{n}(t)-\omega_\bx(t)\right|^ 2\right\}}{\sum_{n=1}^N a_{n}^2(t)}
\end{eqnarray}
where  $a_{n}(t)$ is the amplitude of the $n$th analytic signal, and so forth. The first of these states that amplitude modulation, as well as departures of the instantaneous frequencies from the global mean frequency $\overline\omega_\bx$, contribute to the second instantaneous central moment. The second states that amplitude modulations together with departures of the instantaneous frequencies from the time-varying joint instantaneous frequency $\omega_\bx(t)$ contribute to the squared joint instantaneous bandwidth. In both cases contributions from different times and different signal  components are weighted according to the instantaneous power $a_{n}^2(t)$.

\subsection{Invariance}
The joint instantaneous moments defined above have the important property that they are invariant to transformations of the form
\begin{equation}
\by_+(t)  \equiv  c\bU\bx_+(t) \label{lineartransform}
\end{equation}
where $c$ is some constant and $\bU$ is an $N\times N$ unitary matrix;  these may be termed  \emph{scaled unitary transformations}.  Since in the frequency domain we have also $\bY_+(\omega)  =  c\bU\bX_+(\omega)$, it is obvious that
\begin{equation}
S_\by(\omega)\equiv \frac{\bY_+^H(\omega)\bY_+(\omega)}{\frac{1}{2\pi}\int_{-\infty}^\infty\bY_+^H(\omega)\bY_+(\omega)d \omega}=S_\bx(\omega)
\end{equation}
hence the joint analytic spectrum is unchanged, as are the joint global moments $\overline \omega_{\by}=\overline \omega_{\bx}$ and $\overline \sigma_{\by}^2=\overline \sigma_{\bx}^2$.  The joint instantaneous frequency transforms as
\begin{multline}
\omega_{\by}(t)\equiv \frac{\Im\left\{\by_+^H(t)\frac{d}{d t} \by_+(t)\right\}}{\|\by_+(t)\|^ 2}\\=
\frac{\Im\left\{\bx_+^H(t)\bU^H\bU\,\frac{d}{d t} \bx_+(t)\right\}}{\bx_+^H(t)\,\bx_+(t)}=\omega_{\bx}(t)
\end{multline}
and hence remains unchanged. Similarly it is easy to see that the joint instantaneous bandwidth $\upsilon_{\bx}(t)$ and joint instantaneous second central moments $\sigma_{\bx}^2(t)$ and are also all invariant to scaled unitary transformations.  In particular, these joint instantaneous moments are invariant to coordinate rotations.

\subsection{Joint Bivariate Moments}
We now examine the joint instantaneous moments for bivariate signals by taking either $\bx_+(t)=\begin{bmatrix} z_+(t) & z_-(t)\end{bmatrix}$ or $\bx_+(t)=\begin{bmatrix} x_+(t) & y_+(t)\end{bmatrix}$.  Since these two vectors are related by the scaled unitary transformation (\ref{rotarytoCartesian}), from the invariance shown in the preceding subsection, the results will be identical in either case.  The bivariate joint instantaneous moments integrate to the global moments of the joint analytic spectrum
\begin{equation}
S_z(\omega)\equiv \frac{\left|X_+(\omega)\right|^ 2+\left|Y_+(\omega)\right|^ 2}{ \calE_x+\calE_y}=
\frac{\left|Z_+(\omega)\right|^ 2+\left|Z_-(\omega)\right|^ 2}{\calE_++\calE_-}\label{jointanalytic}
\end{equation}
where $\calE_x\equiv\frac{1}{2\pi}\int_{0}^\infty|X_+(\omega)|^ 2d\omega$, and so forth.  Note that we will use the subscript ``$z$'' rather than the subscript ``$\bx$'' to specifically denote bivariate quantities.

To see (\ref{jointanalytic}), it is helpful to relate the joint analytic spectrum to the two-sided spectrum of the complex-valued signal $z(t)$, which has a Fourier transform $Z(\omega)$.  The Fourier transforms of $z_+(t)$ and $z_-(t)$ are, respectively,
\begin{eqnarray}
Z_+(\omega)&=&U(\omega)Z(\omega)=\frac{1}{2}\left[X_+(\omega)+i Y_+(\omega)\right]\label{rotaryonefrequency}\\
Z_-(\omega)&=&U(\omega)Z(-\omega)=\frac{1}{2}\left[X_+(\omega)-iY_+(\omega)\right]\label{rotarytwofrequency}
\end{eqnarray}
where $Z(\omega)$ is the Fourier transform of  $z(t)$ and where $U(\omega)$ is again the unit step function.  Inserting these into the second and third expressions in (\ref{jointanalytic}), we note at once a cancelation of cross-terms, and the equality follows.  Note that (\ref{rotaryonefrequency}--\ref{rotarytwofrequency}) show that  $z_+(t)$ and $z_-(t)$ have Fourier coefficients drawn entirely from the positive-frequency and negative-frequency sides of $Z(\omega)$, respectively (hence the notation ``+'' and ``-''). The joint analytic spectrum simply consists of averaging the positive and negative frequency halves of $|Z(\omega)|^2$, followed by a normalization to unit energy.

\subsubsection{Instantaneous Frequency}
To obtain the bivariate instantaneous frequency, we begin with  the definition (\ref{multivariatefrequency}) and insert expressions (\ref{plusminusamplitude}) and (\ref{freqgreater}) for the rotary instantaneous frequencies and amplitudes.  This leads to
\begin{equation}
\label{omegaz}
\omega_{z}(t)
=\omega_\phi(t)+ r_z\sqrt{1-\lambda^2(t)}\,\omega_\theta(t)
\end{equation}
for the bivariate instantaneous frequency written in terms of the ellipse parameters. Alternatively, we could have used expressions (\ref{amplitudex}--\ref{amplitudey}) and (\ref{freqx}--\ref{freqy}) from  Appendix~\ref{section:Cartesian} for the Cartesian analytic signals.  Inserting these into (\ref{multivariatefrequency}) we obtain again
(\ref{omegaz}), as we must on account of the invariance proved in the last subsection.  The relative difficulty of directly manipulating the cumbersome Cartesian expressions illustrates the importance of this general result.

The joint bivariate instantaneous frequency $\omega_z(t)$ gives a correct measure of the time-varying frequency content of a bivariate signal regardless of the polarization state.  For $|\lambda(t)|=0$, the signal is purely circular, and $\omega_{z}(t)$ becomes $\omega_\phi(t)+ r_z\omega_\theta(t)$.  Comparison with (\ref{freqgreater}) shows that this is $\omega_+(t)$ if $r_z>0$ and $\omega_-(t)$ if $r_z<0$, that is, for  circularly polarized motion $\omega_{z}(t)$ becomes the instantaneous frequency of the non-vanishing rotary component.  On the other hand, in Appendix~\ref{section:Cartesian} it is shown that for a purely linear signal, $|\lambda(t)|=1$ or vanishing minor axis $|b(t)|=0$, the Cartesian instantaneous frequency along the coordinate axis aligned with the ellipse major axis is equal to $\omega_\phi(t)$, but (\ref{omegaz}) shows that with $|\lambda(t)|=1$ then this is also equal to $\omega_{z}(t)$.  Thus the use of $\omega_z(t)$ is quite desirable as it appropriate for any polarization, unlike the partitioning into a rotary pair or a Cartesian pair.

\subsubsection{Instantaneous Bandwidth}

Similarly the bivariate bandwidth, expressed in terms of the ellipse parameters, is
\begin{equation}
\upsilon_{z}^2(t) = 
\left|\frac{d \ln \kappa(t)}{d t}\right|^ 2 +\frac{1}{1- \lambda^2(t)}\left|\frac{1}{2}\frac{d \lambda(t)}{d t}\right|^ 2+ \lambda^2(t)\omega_\theta^2(t)\label{bivariatebandwidth}
\end{equation}
as follows from the definition (\ref{instantaneousbandwidthexpression}) together with (\ref{bandwidthgreater}) or  (\ref{upsilonx}--\ref{upsilony}).  This expression is quite illuminating.  The first term in (\ref{bivariatebandwidth}) measures the strength of amplitude modulation, the second is the rate of ellipse distortion on account of changing eccentricity, and the third is due to the precession of the ellipse. Note each of the these quantities is independent of coordinate rotation, that is, the orientation angle $\theta(t)$ does not explicitly appear.  We define these three terms as
\begin{equation}
\upsilon_{z}^2(t) \equiv \upsilon_{\kappa}^2(t)+\upsilon_{\lambda}^2(t)+\upsilon_{\theta}^2(t)
\end{equation}
which we call the squared \emph{amplitude bandwidth}, \emph{deformation bandwidth}, and \emph{precession bandwidth}, respectively.

\begin{figure*}[t]
\begin{center}
\includegraphics[height=6in,angle=270]{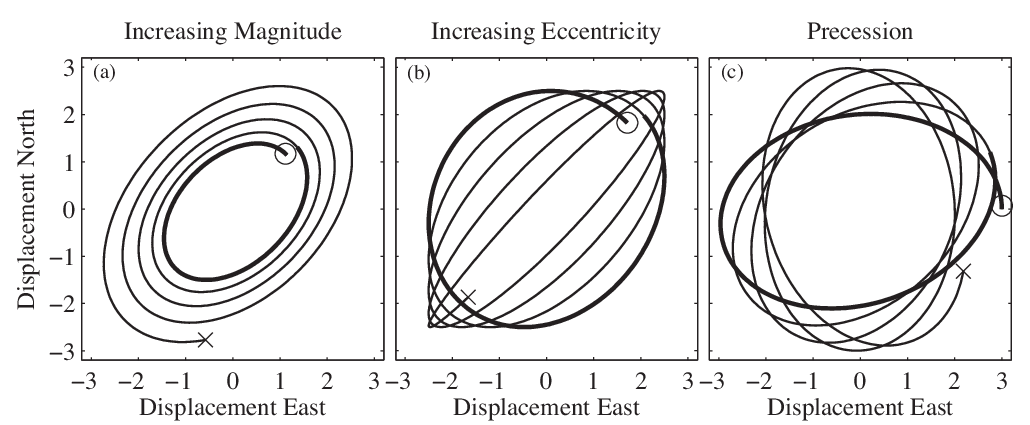}
\end{center}
            \caption{An ellipse with uniformly increasing relative amplitude (a), i.e. $\upsilon_\kappa(t)/\omega_z(t)$ constant, an ellipse with uniformly increasing eccentricity (b), i.e.   $\upsilon_\lambda(t)/\omega_z(t)$  constant, and an a uniformly precessing ellipse (c) with  $\upsilon_\theta(t)/\omega_z(t)$ constant.  The three constants have been chosen to all be the same value of 0.025.  The bold portion of the line in all three panels show an initial single orbit.  A circle marks the beginning of each record and an ``x'' marks the end.
            }\label{stability_schematic}
\end{figure*}

An illustration of the bivariate bandwidth is presented in Figure~\ref{stability_schematic}.  Three different time-varying elliptical signals are plotted, all having constant orbital frequency $\omega_\phi(t)$ and each with exactly one of the three terms in (\ref{bivariatebandwidth}) being nonzero.  The quantities $\upsilon_{\kappa}(t)$, $\upsilon_{\lambda}(t)$, $\upsilon_{\theta}(t)$ are each set equal to the value of 0.025$\times\omega_\phi(t)$ for the three cases respectively.  It is clear that if any of these terms were to become too large, the usefulness of our description of the signal---as an ellipse the properties of which evolve with time---would become questionable. Thus $\upsilon_{z}(t)$ quantifies the lowest-order departure of the bivariate signal from periodic motion tracing out the periphery of a fixed ellipse.

The bivariate bandwidth is therefore a fundamental quantity reflecting the degree of instability of the elliptical motion.  It is remarkable and surprising that the definition (\ref{multivariatefrequency}), which is a power-weighted average of corresponding univariate quantities, should be identical with (\ref{bivariatebandwidth}) which is clearly an expression of the instability of elliptical motion.  That these expressions are equivalent underscores the fact that bandwidth is itself a measure of oscillation stability, and conversely, that the degree of instability of an elliptical signal is interpretable as a bandwidth.

The deformation bandwidth $\upsilon_{\lambda}(t)$ deserves further comment.  Note that this quantity can be expressed in the equivalent forms
\begin{equation}\left|\upsilon_{\lambda}(t)\right|=
\frac{1}{2}\left|\frac{1}{\sqrt{1 -\lambda^ 2(t)}}\frac{d\lambda(t)}{dt}\right|=\frac{1}{2}\left|\frac{1}{\lambda(t)}\frac{d}{dt}\sqrt{1 -\lambda^ 2(t)}\right|
\end{equation}
as may readily be verified.  These lead to the following approximations
\begin{eqnarray}
\left|\upsilon_{\lambda}(t)\right|&=&
\frac{1}{2}\left|\frac{d\lambda(t)}{dt}\right|\left[1+O\left(\lambda^2(t)\right)\right]\\
&=&\frac{1}{2}\left|\frac{d}{dt}\sqrt{1 -\lambda^ 2(t)}\right|\left[1+O\left(1-\lambda^2(t)\right)\right]
\end{eqnarray}
appropriate for the near-circular case $|\lambda(t)|\approx 0$ and the near-linear case $|\lambda(t)|\approx 1$, respectively.  In the near-circular case, the deformation bandwidth is due to the (small) departures of the linearity $|\lambda(t)|$ from zero, while in the near-linear case it is due to the (small) departures of the linearity from unity.  Another interpretation may be found by noting, using (\ref{adef}--\ref{bdef}),
\begin{equation}\left|\upsilon_{\lambda}(t)\right|=
\left|\frac{a(t)b(t)}{a^ 2(t)+b^ 2(t)}\frac{d}{d t} \ln\left[\frac{b(t)}{a(t)}\right]\right|
\end{equation}
which states that the deformation bandwidth is due to fractional changes in the ellipse aspect ratio $b(t)/a(t)$, weighted in proportion to the ratio of the ellipse area $\pi a(t)|b(t)|$ to the root-mean-square radius.

\section{Application}\label{application}
This section presents an application of the bivariate instantaneous frequency and bandwidth to a typical signal from physical oceanography.

\subsection{Data and Method}
As discussed already in the introduction, Fig.~\ref{bandwidth_looper}a presents data from a subsurface, acoustically-tracked oceanographic float \cite{rossby86-jaot}. The data was downloaded from the World Ocean Circulation Experiment Subsurface Float Data Assembly Center (WFDAC) at \url{http://wfdac.whoi.edu}.  This particular record, from an experiment in the eastern North Atlantic Ocean that is well-known among oceanographers \cite{richardson89-jpo,armi89-jpo,spall93-jmr}, was recorded somewhat to the south and west of the Canary Islands. The float was drifting at a depth between 1000 m and 1300 m.  Its horizontal position was recorded once per day, as inferred by triangulation of acoustic travel times from several nearby sound sources.  The region shown ranges from 20--25\degs N and 27--22.5\degs W.

The loops in the trajectory are known to be the imprint of one of the large ``eddies'' or ``vortices'' \cite{mcwilliams85-rvg} that are common in the ocean, a distant relative of the swirls one observes when stirring coffee; note that the loops seen in Fig.~\ref{bandwidth_looper}a are up to fifty kilometers in diameter.  An instrument trapped in such a vortex will record both nearly circular motion from orbiting around the vortex center, which is well modeled as a modulated elliptical signal as shown by \cite{lilly06-npg}, as well as translational or advected motion of the vortex center itself.  This suggests modeling the observed signal $z^{\{o\}}(t)=x^{\{o\}}(t)+iy^{\{o\}}(t)$ via the unobserved components model \cite{harvey89}
\begin{equation}
z^{\{o\}}(t)=z(t)+z^{\{\epsilon\}}(t)
\end{equation}
where $z(t)$ is a modulated elliptical signal and where $z^{\{\epsilon\}}(t)$ is a residual defined to include everything else.  The residual $z^{\{\epsilon\}}(t)$ is expected to include the turbulent background flow together some measurement noise, but may be considered ``noise'' at present since we are interested only in $z(t)$.

A means of estimating $z(t)$ using an extension of wavelet ridge analysis \cite{delprat92-itit,mallat} was developed by \cite{lilly06-npg}.  The details of this method are not particularly important here, except to note that it is expected to give a reasonable estimate $\widehat z(t)$ of the presumed (unobserved) modulated elliptical signal $z(t)$.  We apply this method, with parameter choices noted in Appendix~\ref{section:numericaldetails}, to the record in Fig.~\ref{bandwidth_looper}a.  This leads to the estimated bivariate oscillation $\widehat z(t)$ shown in  Fig.~\ref{bandwidth_looper}b, and---by subtraction, $\widehat z^{\{\epsilon\}}(t)\equiv z^{\{o\}}(t)-\widehat z(t)$---to the estimated residual signal $\widehat z^{\{\epsilon\}}(t)$ shown in Fig.~\ref{bandwidth_looper}c.  In Fig.~\ref{bandwidth_looper}b the signal $\widehat z(t)$ is represented by snapshots of the modulated signal at different times, by letting the orbital phase vary with the ellipse geometry held fixed, as described earlier in the discussion of (\ref{ellipseeqnseparated}). Note that in Fig.~\ref{bandwidth_looper}(c) virtually all the looping motions have been removed, leaving behind a large-scale meander plus small-scale irregularities.

\subsection{Instantaneous Amplitude and Frequency}

\begin{figure}[t]
\begin{center}
\hspace{-0.2cm}\includegraphics[width=3.5in,angle=0]{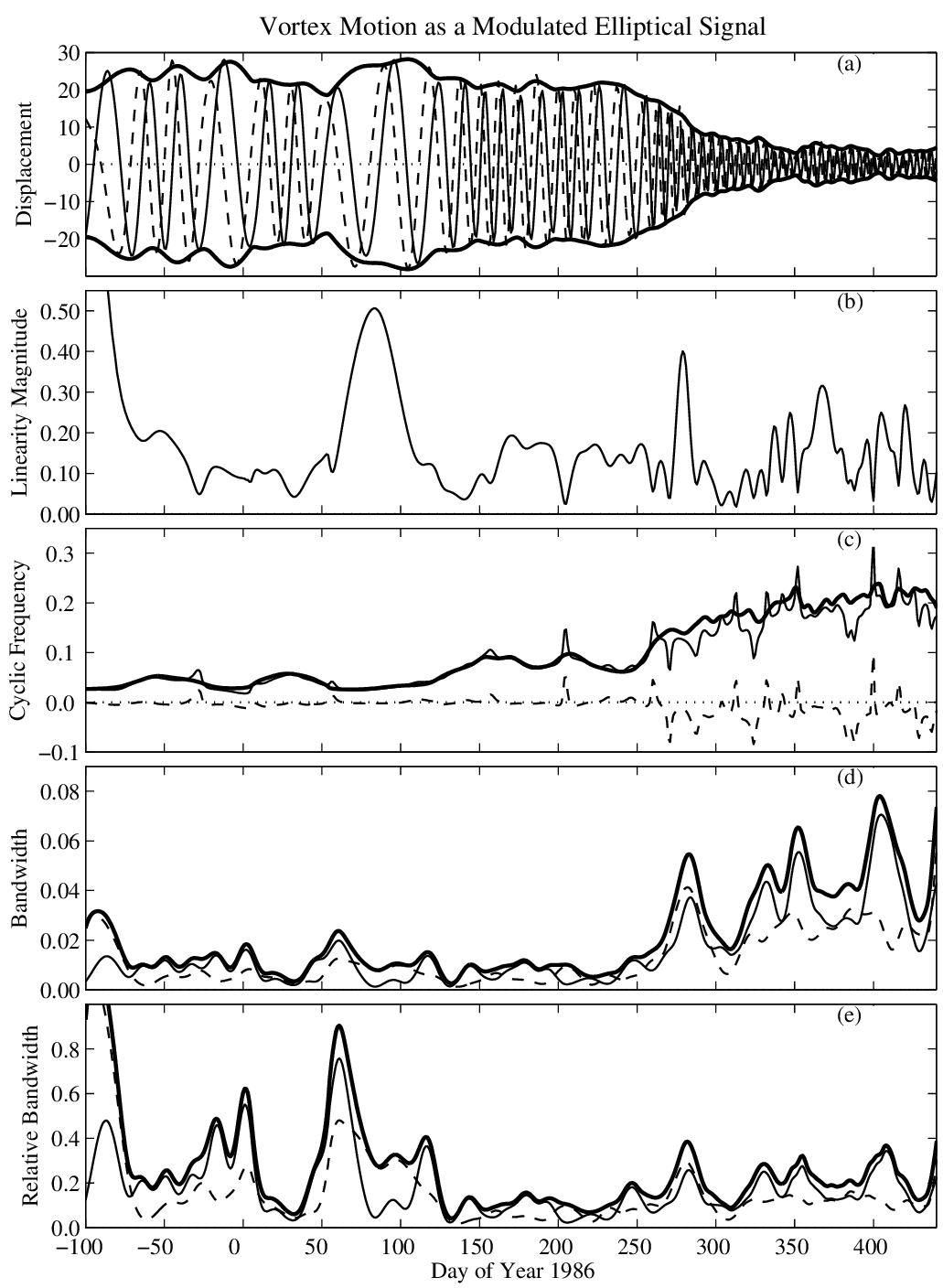}
\end{center}
            \caption{Instantaneous moment analysis of the estimated modulated elliptical signal $\widehat z(t)$, derived from the bivariate time series of Fig.~\ref{bandwidth_looper}a as described in the text.  Panel (a) shows the real and imaginary parts of the signal $\widehat z(t)$, $\widehat x(t)$ (thin solid line) and $\widehat y(t)$ (dashed line), together with the associated RMS amplitude $\widehat \kappa(t)$ (heavy solid line).   In panel (b), the linearity $|\widehat \lambda(t)|$ is shown.  Three time-varying frequencies are shown in panel (c), the joint instantaneous frequency $\widehat \omega_z(t)$ (heavy solid line), the orbital frequency $\widehat\omega_\phi(t)$ (thin solid line), and the precession rate $\widehat \omega_\theta(t)$ (thin dashed line). Panel (d) presents the three terms in the bivariate instantaneous bandwidth, the amplitude bandwidth $\widehat \upsilon_\kappa(t)$ (heavy solid line), the deformation bandwidth $\widehat \upsilon_\lambda(t)$ (thin solid line), and the precession bandwidth $\widehat \upsilon_\theta(t)$ (dashed line); in (e), the bandwidths are presented again, but this time divided through by $|\widehat \omega_z(t)|$ in order to render them nondimensional.
The units of (a) are kilometers, the units of (c) and (d) are radians per day, and  (b) and (e) are non-dimensional and so have no units.
            }\label{bandwidth_signal}
\end{figure}

Having achieved this decomposition, we now analyze in more detail the properties of the estimated bivariate oscillatory signal $\widehat z(t)$ using the joint instantaneous moments. The estimated signal, shown in Fig.~\ref{bandwidth_signal}a together with the instantaneous root-mean-square amplitude $\widehat \kappa(t)$, exhibits both substantial amplitude as well as frequency modulations; note that we will denote all properties of $\widehat z(t)$ with a hat, `` $\widehat\cdot$ '',  to distinguish them from the properties of the unobserved true signal $z(t)$.  In particular, during the last third of the record, say after yearday 250, the signal amplitude is greatly reduced compared to the earlier time period.  This accounts for the transition from large to small ellipses seen in Fig.~\ref{bandwidth_looper}b.

This signal rotates in a clockwise fashion: the negative rotary component $\widehat z_-(t)$ dominates the positive component $\widehat z_+(t)$, thus the signed linearity $\widehat \lambda(t)$ (not shown) is everywhere negative; this sense of rotation can be inferred from the phase shift between the real and imaginary parts of $\widehat z(t)$ in Fig.~\ref{bandwidth_signal}a.  The linearity $|\widehat \lambda(t)|$ itself (Fig.~\ref{bandwidth_signal}b) is generally very small, corresponding to nearly circular motion, apart from a few excursions to higher values. Thus only a handful of the ellipses shown in Fig.~\ref{bandwidth_looper}b exhibit substantial eccentricity.

The instantaneous frequency content of the signal is shown Fig.~\ref{bandwidth_signal}c. Here, the bivariate instantaneous frequency of $\widehat z(t)$, $\widehat\omega_z(t)$, is presented together with its orbital frequency $\widehat\omega_\phi(t)$ and precession rate $\widehat\omega_\theta(t)$.  The precession rate is seen to be much smaller than the orbital frequency, and consequently, it follows from (\ref{omegaz}) that the bivariate instantaneous frequency $\widehat \omega_z(t)$ should be close to the orbital frequency $\widehat \omega_\phi(t)$.  A transition is seen around yearday 250 to higher-frequency oscillations, corresponding to the transition to smaller amplitudes  seen in Fig.~\ref{bandwidth_signal}a.  Interestingly, in the higher-frequency portion of the record, the precession rate exhibits a tendency for negative rotation, but establishing whether or not this result is statistically significant is beyond the scope of the present paper.  The precession rate $\widehat \omega_\theta(t)$ and orbital frequency $\widehat \omega_\phi(t)$ both present some jagged or irregular variability, which is likely due to the effect of measurement noise; this interpretation is supported by the fact that the irregular variability is particularly present after yearday 250, when the signal strength has weakened.

Measurement noise is expected to strongly affect $\widehat \omega_\phi(t)$ and $\widehat \omega_\theta(t)$ in a case such as this one, when then unobserved signal of interest $z(t)$ is nearly circular in nature.  Here we have $|z_-(t)|\gg|z_+(t)|$, and so the observed (noisy) signal $z^{\{o\}}(t)$ will have a much lower signal-to-noise ratio at positive frequencies than at negative frequencies.  The estimated positive rotary signal $\widehat z_+(t)$ is thus expected to be considerably noisier than the negative rotary signal $\widehat z_-(t)$, and likewise for the associated instantaneous frequencies $\widehat \omega_+(t)$ compared with $\widehat \omega_-(t)$. Since $\widehat \omega_\phi(t)$ and $\widehat \omega_\theta(t)$ are, respectively, the sum and difference of  $\widehat \omega_+(t)$  and $\widehat \omega_-(t)$, noise in $\widehat \omega_+(t)$ will affect them both.

By contrast $\widehat \omega_z(t)$ is an average over the instantaneous frequencies of two independent signal components. It is therefore expected to be more robust against the effects of noise than $\widehat \omega_+(t)$ and  $\widehat \omega_-(t)$, and also than the Cartesian analytic signals $\widehat \omega_x(t)$  and  $\widehat \omega_y(t)$ that would be observed along the coordinate axes.  This suggests that although $\widehat \omega_\phi(t)$ and $\widehat \omega_\theta(t)$ are of great interest from the point of view of the modulated ellipse model, $\widehat \omega_z(t)$ provides a superior estimate of the time-varying frequency content of a noisy bivariate oscillatory signal.  The preceding discussion has also brought up the importance of understanding the effects of noise on the estimation procedure, a task which is currently underway.

\subsection{Instantaneous Bandwidth}

The three terms in the bivariate instantaneous bandwidth of the estimated signal $\widehat z(t)$---$\widehat \upsilon_\kappa(t)$,  $\widehat \upsilon_\lambda(t)$, and $\widehat \upsilon_\theta(t)$---are shown in Fig.~\ref{bandwidth_signal}d. During most of the record, the amplitude bandwidth $\widehat \upsilon_\kappa(t)$ is larger than the other two. Since these contribute as their squares to the bivariate bandwidth $\widehat \upsilon_z(t)$, we see that in fact the amplitude bandwidth $\widehat \upsilon_\kappa(t)$ accounts for the large majority of the estimated joint instantaneous bandwidth $\widehat \upsilon_z(t)$ during most of the record. Thus the lowest-order departure of the bivariate signal from a uniform oscillation at frequency $\widehat \omega_z(t)$ is generally due to the modulation of the root-mean-square ellipse amplitude, with its eccentricity and orientation held fixed; this point will be discussed further in the next subsection.

The three bandwidth quantities are again shown in Fig.~\ref{bandwidth_signal}e, but this time each has been divided by $\widehat \omega_z(t)$, thus comparing the contributions to the instantaneous bandwidth with the instantaneous frequency.  Whereas Fig.~\ref{bandwidth_signal}d suggests that the instantaneous bandwidth increases after yearday 250, Fig.~\ref{bandwidth_signal}e shows that this transition is offset by the corresponding increase in instantaneous frequency.  This means that the variability of the modulated elliptical, on time scales proportional to the local period of oscillation, is in fact similar during the different periods of the record. If anything, the ``relative bandwidth'' is somewhat lower during the last third of the record then in the remainder.

\subsection{Interpretation}
The preceding results have interesting physical interpretations.  The observed signals primarily correspond to the motion of a water parcel trapped in a vortex, but it is also known that the floats tend to behave slightly differently than actual fluid does; see the discussion in \cite{lilly06-npg} for example.  Thus, the observed evolution of the elliptical signal from low to high frequency, and from large amplitude to small amplitude, could either reflect (i) evolution of the vortex itself or (ii) motion of the float closer to the center of a vortex having more-or-less fixed properties.  While further investigation must be left to the future, it is clear that examining the evolution of the ellipse properties in float trajectories provides a powerful tool for understanding the life cycles of oceanic vortices.

The fact that the vortex motion is apparently not purely circular is by itself interesting.  Most dynamical solutions for vortices of this type, for example \cite{mcwilliams88-jpo}, assume circular symmetry.  However it is also well known  \cite{ruddick87-jpo} that the application of an exterior strain field to a circular vortex can deform it into an ellipse, as the vortex exchanges energy with the ambient flow.  The changing ellipse eccentricity seen in Fig.~\ref{bandwidth_signal}b could therefore indicate a modulation of the vortex shape by a large-scale strain field. To date, vortex eccentricity has rarely been examined from oceanographic observations, owing in large part to the lack of a suitable analysis method.  It should be emphasized that any interpretation of eccentricity must carefully treat the statistics of the noise component of the signal, since noise cannot make a circular vortex appear more circular.  

The three contributions to the bivariate bandwidth, presented earlier in Fig.~\ref{stability_schematic}, each correspond to different physical pathways in the evolution of a vortex.  Increasing vortex magnitude could indicate so-called balanced adjustment, in which the vortex radius and hence thickness change but key integral properties remain fixed \cite{mcwilliams88-jpo}.  Increasing eccentricity with a fixed orientation could indicate deformation of a vortex in a strain field, as discussed above.  Finally, it is known that there are also solutions to the equations of motion in which an initially elliptical vortex may evolve by precessing (i.e. changing its orientation) in the absence of exterior forcing \cite{young86-jfm}.  

These three possibilities of adjustment, deformation, and free precession represent very different evolutionary pathways, yet all contribute identically to the spectral bandwidth of a float record.  Thus the important physical differences are obscured if one only examines global spectral quantities.  In this case, we have seen that the instantaneous bandwidth is dominated by the amplitude bandwidth.  Whether this corresponds physically to vortex adjustment, or instead is an artifact of possible float motion within the vortex, is an intriguing and important question for further research.

\section{Discussion}

This paper has examined the extension of instantaneous moments to bivariate signals.    To accomplish this, ``joint instantaneous moments'' were defined for a general multivariate signal.  These  are time-varying quantities that integrate to the global moments of the average analytic spectrum, that is, the spectrum of the analytic part of the signal components averaged over the number of components.  While the joint instantaneous frequency is simply the power-weighted average of the instantaneous frequencies of the components, the joint instantaneous bandwidth has an unexpected but very intuitive form.  It measures the extent to which a multivariate oscillation does not evolve simply by oscillating at the single, time-varying frequency given by the joint instantaneous frequency.

These joint instantaneous moments together with the notion of a modulated elliptical signal represent a powerful means for analyzing bivariate signals.  The ``modulated elliptical signal'' is a representation of a bivariate oscillation, alternately regarded as a \emph{pair} of analytic signals, as a single time-varying structure.  The joint moments describe the nature of the variability of this structure, combining the partial descriptions achieved from the individual moments of the members of a signal pair.  Expressing the bivariate instantaneous frequency and bandwidth in terms of the time-varying ellipse parameters leads to illuminating expressions.  In particular, the instantaneous bandwidth was shown to consist of three terms: amplitude modulation, deformation or eccentricity modulation, and orientation modulation or precession.  These three quantities thus express the three basic ways ellipse geometry can change with time; the bivariate bandwidth is therefore seen as the degree of instability of a modulated elliptical signal.

An application of the instantaneous moments to a bivariate time series from oceanography leads to a wealth of information regarding the time variability of the signal, which in this case could reflect time variations in an underlying oceanic vortex structure. In addition to pursing physical questions raised here, the most important outstanding tasks involve quantifying the errors involved in estimating bivariate oscillatory signals.  There are both ``deterministic'' errors due to the time variability of the signal, which have recently been examined for univariate wavelet ridge analysis \cite{lilly09-itit}, as well as random errors due to background noise.  These topics are the subjects of ongoing research.

\appendices

\section{The Cartesian Analytic Signals}\label{section:Cartesian}

Here we derive expressions for the Cartesian instantaneous moments, that is, the instantaneous moments of $x_+(t)=a_x(t) \,e^{i\phi_x(t)}$ and $y_+(t)=a_y(t) \,e^{i\phi_y(t)}$ for a bivariate signal represented as $z(t)=\Re\{x_+(t)\}+i\Re\{y_+(t)\}$.  In terms of the parameters of the modulated elliptical signal of Section~\ref{section:oscillations}, the amplitudes and phases of the Cartesian analytic signals are
\begin{eqnarray}
a_x(t)&= &\kappa(t)\sqrt{1 + |\lambda(t)\,|\cos 2 \theta(t)}\label{amplitudex} \\
a_y(t)&= &\kappa(t)\sqrt{1 - |\lambda(t)\,|\cos 2 \theta(t)}\label{amplitudey} \\
\phi_x(t)&= &
\phi(t)\nonumber\\
&&  \hspace{-0.3in}+\Im\left(\ln\left\{a(t)\cos\theta(t)+i r_z\left|b(t)\right|\sin\theta(t)\right\}\right)\label{phix}\\
\phi_y(t)&= &\phi(t) -r_z\pi/2\nonumber\\
&&  \hspace{-0.3in}+\Im\left(\ln\left\{\left|b(t)\right|\cos \theta(t)+ir_z a(t)\sin \theta(t)\right\}\right)\label{phiy}
\end{eqnarray}
which simplifies the presentation of \cite{lilly06-npg}.  Note that the combination $\Im\left(\ln\left\{z\right\}\right)$ is used to implement the four-quadrant inverse tangent function,  with $\tan\left(\Im\left(\ln\left\{z\right\}\right)\right)=\Im(z)/\Re(z)$. As $|\lambda(t)|$ approaches zero, and the signal is nearly circular, both the Cartesian amplitudes $a_x(t)$ and $a_y(t)$ approach $\kappa(t)$.  Meanwhile the phases $\phi_x(t)$ and $[\phi_y(t)+r_z\pi/2]$ both approach  $\phi(t)+r_z\theta(t)$, which has been previously identified as the phase of whichever rotary component, $z_+(t)$ or $z_-(t)$, has the larger amplitude.

Differentiating the phase expressions (\ref{phix}--\ref{phiy}) one obtains the following, rather complicated, forms for the Cartesian instantaneous frequencies
\begin{eqnarray}
\omega_{x}(t)& \equiv & \frac{d\phi_x(t)}{dt}=\omega_\phi(t)+r_z\frac{\kappa^ 2(t)}{a_x ^ 2(t)}\times\nonumber\\
& & \!\!\!\!\!\!\!\!\!\!\!\!\!\!\!\!\!\!\!\!\!\!\!\!\!\!\left[
\omega_\theta(t)\sqrt{1 -\lambda ^ 2(t)}-\frac{1}{2}\frac{\sin 2\theta(t)}{\sqrt{1 -\lambda ^ 2(t)}}\frac{d|\lambda(t)|}{d t}\right]
\label{freqx}\\
\omega_{y}(t)&\equiv & \frac{d\phi_y(t)}{dt}=\omega_\phi(t)+r_z\frac{\kappa^ 2(t)}{a_y ^ 2(t)}\times  \nonumber\\
& & \!\!\!\!\!\!\!\!\!\!\!\!\!\!\!\!\!\!\!\!\!\!\!\!\!\!\left[
\omega_\theta(t)\sqrt{1 -\lambda ^ 2(t)}+\frac{1}{2}\frac{\sin 2\theta(t)}{\sqrt{1 -\lambda ^ 2(t)}}\frac{d|\lambda(t)|}{d t}\right].
\label{freqy}
\end{eqnarray}
To derive these, we first note that the derivatives of (\ref{phix}--\ref{phiy}) become
\begin{eqnarray}
\!\!\!\!\!\!\!\omega_{x}(t)& \seq & \omega_\phi(t)+\Im\left\{\frac{d}{dt}\ln\left(1+i\frac{b(t)}{a(t)}\tan\theta(t)\right)\right\}
\\\!\!\!\!\!\!\!\omega_{y}(t)& \seq & \omega_\phi(t)+\Im\left\{\frac{d}{dt}\ln\left(1+i\frac{a(t)}{b(t)}\tan\theta(t)\right)\right\}
\end{eqnarray}
after pulling out the real parts of the quantities inside the natural logarithms.  Then using
\begin{multline}
\frac{d}{dt}\ln\left(\frac{b(t)}{a(t)}\right)=
\frac{d}{dt}\ln\left(\frac{\sqrt{1-|\lambda(t)|}}{\sqrt{1+|\lambda(t)|}}\right)=\\
-\frac{1}{1-\lambda^2(t)}\frac{d|\lambda(t)|}{dt}=-\frac{d}{dt}\ln\left(\frac{a(t)}{b(t)}\right)
\end{multline}
we find (\ref{freqx}--\ref{freqy}) in a few straightforward lines of algebra.

The Cartesian frequencies have three terms: the orbital frequency $\omega_\phi(t)$, a weighted version of the precession rate $\omega_\theta(t)$, and an orientation-dependent term involving the deformation rate.  The last term makes an exactly opposite contribution to the two power-weighted instantaneous frequencies $a_x^2(t)\omega_x(t)$ and $a_y^2(t)\omega_y(t)$.  Owing to cancelation of these orientation-dependent terms, we readily obtain (\ref{omegaz}) for the bivariate instantaneous frequency using the definition (\ref{multivariatefrequency}) applied to the Cartesian analytic signal vector $\begin{bmatrix} x_+(t) & y_+(t)\end{bmatrix}^T$.

Now for the instantaneous bandwidths, differentiating the logarithms of (\ref{amplitudex}--\ref{amplitudey}) gives
\begin{eqnarray}
\upsilon_x(t)
& =& \frac{d\ln \kappa(t)}{d t}+
\frac{1}{2}\frac{\kappa^2(t)}{a_x^2(t)}
   \frac{d}{d t}\left[\left|\lambda(t)\right|\cos 2\theta(t)\right]\label{upsilonx}\\
\upsilon_y(t)
&=& \frac{d\ln\kappa(t) }{d t}-\frac{1}{2}\frac{\kappa^2(t)}{a_y^2(t)}
   \frac{d}{d t}\left[\left|\lambda(t)\right|\cos 2\theta(t)\right]\label{upsilony}
\end{eqnarray}
as expressions for Cartesian bandwidths $\upsilon_x(t) \equiv \frac{d}{dt}\ln a_x(t)$ and $\upsilon_y(t) \equiv \frac{d}{dt}\ln a_y(t)$, respectively.  In both of these there are three terms, due to amplitude modulation, precession, and deformation, respectively; the contributions from each of the latter two terms are orientation-dependent. These expressions together with the multivariate instantaneous bandwidth definition (\ref{instantaneousbandwidthexpression}) give (\ref{bivariatebandwidth}) for the form of the bivariate bandwidth.

\section{Numerical Method}\label{section:numericaldetails}
This appendix describes the means of generating the estimated elliptical signal $\widehat z(t)$ shown in Fig.~\ref{bandwidth_looper}b and Fig.~\ref{bandwidth_signal}. The reader is asked to refer to \cite{lilly06-npg} for details; here we only give a brief overview together with parameter settings.  The wavelet transform using a generalized Morse wavelet \cite{olhede02-itsp,lilly09a-itsp} with parameter choices $\beta=3$ and $\gamma=3$ is performed on two real value time  series $x(t)$ and $y(t)$.  This transform is performed at fifty logarithmically spaced frequency bands ranging from one cycle per 53 days to one cycle per 2.6 days.

Wavelet ridge analysis consists of locating a maximum of the transform modulus across scale at each time, then connecting these points across time into a continuous chain \cite{delprat92-itit,mallat,lilly09-itit}.  The wavelet transform along this so-called ``ridge curve'' constitutes an estimate of a (univariate) modulated oscillatory signal. It was shown by \cite{lilly06-npg} that way the wavelet ridge curves of the $x(t)$ and $y(t)$ time series, constructed separately, can be combined to form estimates of a modulated elliptical, or bivariate oscillatory, signal.  Application of this method to the time series shown in Fig.~\ref{bandwidth_looper}a gives the estimated elliptical signal $\widehat z(t)$.

\section*{Acknowledgment}
We thank two anonymous reviewers for helpful comments.

\begin{biography}{Jonathan M. Lilly}
(M05) was born in Lansing, MI,
in 1972. He received the B.S. degree in geology and
geophysics from Yale University, New Haven, CT,
in 1994, and the M.S. and Ph.D. degrees in physical
oceanography from the University of Washington,
Seattle, in 1997 and 2002, respectively.
He was a Postdoctoral Researcher with the
Applied Physics Laboratory and School of Oceanography,
University of Washington, from 2002 to
2003, and with the Laboratoire d'Oc\'eanographie
Dynamique et de Climatologie, Universit\'e Pierre
et Marie Curie, Paris, France, from 2003 to 2005. Since 2005, he has been
a Research Associate with Earth and Space Research, a nonprofit scientific
institute in Seattle. His research interests are oceanic vortex structures, satellite
oceanography, time/frequency analysis methods, and wave--wave interactions.
Dr. Lilly is a member of the American Meteorological Society and of the
American Geophysical Union.
\end{biography}

\begin{biography}{Sofia C. Olhede}
was born in Spanga, Sweden, in 1977. She received the M. Sci. and
Ph.D. degrees in mathematics from Imperial College
London, London, U.K., in 2000 and 2003,
respectively.
She held the posts of Lecturer (2002-2006) and
Senior Lecturer (2006-2007) with the Mathematics
Department, Imperial College London, and in 2007,
she joined the Department of Statistical Science, University
College London, where she is Professor of
Statistics.
Her research interests include the analysis of complex-valued stochastic processes, non-stationary time series and inhomogeneous random fields.
Prof. Olhede serves on the Research Section of the Royal Statistical Society and is an Associate Editor of the Journal of the Royal Statistical
Society, Series B (Statistical Methodology). 
\end{biography}

\label{lastpage}

\end{document}